\documentclass[preprint,showpacs,preprintnumbers,amsmath,amssymb,nofootinbib,showkeys,aps]{revtex4}
\usepackage{epsfig}

\usepackage{graphicx}
\usepackage{dcolumn}
\usepackage{bm}
\usepackage{amsmath}
\usepackage{amssymb}
\usepackage{latexsym}
\usepackage{color}
\usepackage{hyperref}
\usepackage{mathrsfs}

\newcommand{\be}{\begin{equation}}
\newcommand{\ee}{\end{equation}}
\newcommand{\bea}{\begin{eqnarray}}
\newcommand{\eea}{\end{eqnarray}}

\newcommand{\like}{\mathscr{L}}
\begin{document}


\title{Gaussian Process Estimation of Transition Redshift}

\author{J. F. Jesus$^{1,2}$}\email{jf.jesus@unesp.br}
\author{R. Valentim$^{3}$} \email{valentim.rodolfo@unifesp.br}
\author{A. A. Escobal$^{2}$} \email{anderson.aescobal@gmail.com}
\author{S. H. Pereira$^{2}$} \email{s.pereira@unesp.br}

\affiliation{$^1$Universidade Estadual Paulista (UNESP), Campus Experimental de Itapeva - R. Geraldo Alckmin, 519, 18409-010, Itapeva, SP, Brazil,
\\$^2$Universidade Estadual Paulista (UNESP), Faculdade de Engenharia de Guaratinguet\'a, Departamento de F\'isica e Qu\'imica - Av. Dr. Ariberto Pereira da Cunha 333, 12516-410, Guaratinguet\'a, SP, Brazil
\\$^3$Universidade Federal de S\~ao Paulo (UNIFESP), Departamento de F\'{\i}sica, Instituto de Ci\^encias Ambientais, Qu\'{\i}micas e Farmac\^euticas (ICAQF), Rua S\~ao Nicolau 210, 09913-030, Diadema, SP, Brazil}


\def\zt{\mbox{$z_t$}}

\begin{abstract}
This paper aims to put constraints on the transition redshift $z_t$, which determines the onset of cosmic acceleration, in  cosmological-model independent frameworks. In order to do that, we use the non-parametric Gaussian Process method with $H(z)$ and SNe Ia data. The deceleration parameter reconstruction from $H(z)$ data yields $z_t=0.59^{+0.12}_{-0.11}$. The reconstruction from SNe Ia data assumes spatial flatness and yields $z_t=0.683^{+0.11}_{-0.082}$. These results were found with a Gaussian kernel and we show that they are consistent with two other kernel choices.


\end{abstract}

\maketitle



\section{Introduction} 
It is well known that we live in a special phase of accelerated expansion of universe, as indicated by SNe Ia\footnote{Type Ia Supernovae} observations \cite{SN1,SN2,SN3,SN4,SN5,union,union2,union21} and also by other complementary observations such as CMB\footnote{Cosmic Microwave Background} radiation \cite{WMAP1,WMAP2,planck}, BAO\footnote{Baryonic Acoustic Oscillations} \cite{BAO1,BAO2,BAO3,BAO4,BAO5} and $H(z)$\footnote{Hubble parameter} measurements \cite{Omer,Omer2,Omer3}. From a theoretical perspective, the so-called $\Lambda$CDM model accommodates quite well such accelerating phase, with free parameters being accurately constrained \cite{planck,Omer2,sharov}.

Determining the exact moment in the history of evolution of the universe in which accelerated phase began is an interesting question, since different models should provide the same result. The parameter that determines such transition from deceleration to an accelerated phase is called transition redshift, $z_t$, which can be treated as a new cosmic parameter \cite{Omer2,Omer3,kine3,cunha2008,guimaraes2009,Rani2015, limajesus2014,moresco2016,jrs2018}. A model without a reasonable transition redshift, $z_t\sim0.5-1$, for instance, fails on explaining current cosmological observations. Thus, currently, the transition redshift $z_t$ has similar importance as the deceleration parameter $q_0$ and Hubble constant $H_0$.

Different approaches have been used to constrain the transition redshift. Some of them, sometimes called model independent (or kinematic) approaches, use a parametrization for the deceleration parameter as a function of the redshift, $q(z)$, and the values for $z_t$ can be obtained depending on the fixed parameters constrained to $q(z)$. For instance, in a linear parameterization of the form $q(z)=q_0+q_1 z$, the value for $z_t$ found in the literature was: $z_t=0.43^{+0.09}_{-0.05}$ from 182 SNe Ia \cite{SN4} \cite{cunha2008}, $z_t = 0.61^{+3.68}_{-0.21}$ from SNLS \cite{SN3} and $z_t = 0.60^{+0.28}_{-0.11}$ from Davis {\it et al} \cite{SN5}. Reference \cite{guimaraes2009}  found $z_t=0.49^{+0.27}_{-0.09}$ from Union compilation \cite{union} and \cite{Rani2015} found $z_t \approx 0.98$ from a combined analysis of ages, lensing and SNe Ia data. For a linear parameterization on the scale factor, $q=q_0+q_1(1-a)=q_0+q_1 z/(1+z)$, Union+BAO+$H(z)$ data yielded $z_t=0.609^{+0.110}_{-0.070}$ \cite{Xu2009}. By using an $H(z)$ parameterization which tried to highlight the transition redshift and the results from BOSS\footnote{Baryon Oscillation Spectroscopic Survey \cite{BAO3,BAO4,BAO5}}, reference \cite{moresco2016} have obtained a kinematic determination of the transition redshift as $z_t = 0.4 \pm 0.1$ (see also \cite{Omer,Omer2,Omer3}). Recently, a study \cite{jrs2018} of the transition redshift by means of polynomial parametrizations of the comoving distance, $H(z)$ and $q(z)$, found $z_t=0.806\pm 0.094$, $z_t=0.870\pm 0.063$ and $z_t=0.973\pm 0.058$ at 1$\sigma$ c.l., respectively for each parametrization from a combination of SNe Ia \cite{JLA} and $H(z)$ data. The above cited methods are called parametric methods, since a priori parametrization on the background cosmological functions are supposed and then the free coefficients are constrained to observational data.

A much more interesting method to access the cosmological parameters has been proposed recently by Seikel, Clarkson and Smith \cite{seikel2012}, the so called Gaussian Process (GP). In such approach an statistical, non-parametric method is used to reconstruct the dependence with redshift of a cosmological observable, which can be the expansion rate, luminosity distance etc. This Bayesian approach consists in a generalisation of the concept of distributions, describing a distribution
over functions, with a prior, likelihood of observational data, and corresponding posterior function distribution. 

With the aid of GP method, one can reconstruct a general function directly from data, without the need for assume a particular parameterization for it \cite{Rasmussen}. Seikel et al. \cite{seikel2012} has used GP to reconstruct the luminosity distance, $d_L(z)$, and its derivatives, which, combined, allows to reconstruct the dark energy equation of state parameter $\omega(z)$. The reconstruction of the dark energy evolution at low and moderate redshift is quite well. Holsclaw et al. \cite{Hols1,Hols2} use GP in combination with Monte Carlo methods to reconstruct directly the equation of state parameter. In Ref. \cite{Shafieloo} GP is used to constrain the Hubble parameter and deceleration parameter as a function of redshift. The non-parametric reconstruction of the growth index from observational data by GP method was recently done in \cite{Yin} and cosmological tests with strong gravitational lenses was done in \cite{Yenn1}. A reconstruction of the HII galaxy Hubble diagram using GP was done in \cite{Yenn2}. Six different cosmological models were studied in \cite{Yenn3} with GP, showing a preference for a $R_h=ct$ type universe \cite{melia2012}.

In the present paper we reconstruct the deceleration parameter from $H(z)$ data and SNe Ia and then the transition redshift $z_t$ is obtained by using GP method. It was done the reconstruction of $H(z)$ and its derivative together with their uncertainties, and then the transition redshift was obtained from $q(z)$. The same procedure was done with SNe Ia, where we used the apparent magnitude data to estimate luminosity distances, reconstructed the $d_L(z)$ function and its derivatives and then have obtained $q(z)$ reconstruction. From $q(z)$ reconstructions we have obtained transition redshift estimates.

The paper is organized as follows. In Section \ref{GP} we present the main equations of Gaussian Processes. In Section \ref{basic} we present the cosmological equations which relate $z_t$ to the cosmological observables, luminosity distance and $H(z)$. Section \ref{samples} presents the $H(z)$ and SNe Ia data set used and the analyses are presented in Section \ref{analysis}. Conclusions are left to Section \ref{conclusion}. 

\section{\label{GP}Gaussian Processes}
The goal of the Gaussian Process method is to reconstruct a function $f(x)$ from a set of its measured values $f(x_i)\pm\sigma_i$. It assumes that the value of the function at any point $x$ follows a Gaussian distribution. The value of the function at $x$ is correlated with the value at other point $x'$. That is why a covariance (or kernel) $k(x,x')$ function is needed to estimate the expectation and standard deviation of this distribution from data. The distribution of functions is then described by
\be
\mu(x)=<f(x)>,\quad k(x,x')=<(f(x)-\mu(x))(f(x')-\mu(x'))>,\quad\mathrm{Var}(x)=k(x,x)
\ee

We may write the Gaussian Process as
\be
f(x)\sim\mathcal{GP}(\mu(x),k(x,x'))
\ee
The GP method is a non-parametric method because it does not depend on a set of model free parameters to be constrained, but it depends on the choice of covariance function. The covariance function in general depends on the distance between the input points $|x-x'|$ and the covariance is expected to be larger when the input points are close to each other. Here, we consider three covariance functions in order to test this dependence of our results. The usual covariance function, most used in the analyses in the literature is the Squared Exponential (or Gaussian):
\be
k(x,x')=\sigma_f^2\exp\left[-\frac{(x-x')^2}{2l^2}\right]
\ee
where $\sigma_f$ and $l$ are the so called GP hyperparameters, which control the strength of the correlation of the function value and the coherence length of the correlation in $x$, respectively. The other two covariance functions that we choose to analyse are Mat\'ern(5/2):
\be
k(x,x')=\sigma_f^2\exp\left[-\frac{\sqrt{5}|x-x'|}{l}\right]\left[1+\frac{\sqrt{5}|x-x'|}{l}+\frac{5(x-x')^2}{3l^2}\right]
\ee
and Mat\'ern(7/2):
\be
k(x,x')=\sigma_f^2\exp\left[-\frac{\sqrt{7}|x-x'|}{l}\right]\left[1+\frac{\sqrt{7}|x-x'|}{l}+\frac{14(x-x')^2}{5l^2}+\frac{7\sqrt{7}|x-x'|^3}{15l^3}\right].
\ee

The hyperparameters $\sigma_f$ and $l$ are optimized for the observed data, by minimizing a log marginal likelihood function \cite{seikel2012}:
\bea
\ln\mathcal{L}&=&\ln p(\bm{y}|\bm{X},\sigma_f,l)=\nonumber\\
&=&-\frac{1}{2}(\bm{y}-\bm{\mu})^T[K(\bm{X},\bm{X})+C]^{-1}(\bm{y}-\bm{\mu})-\frac{1}{2}\ln|K(\bm{X},\bm{X})+C|-\frac{n}{2}\ln2\pi
\eea
where $\bm{X}={x_i}$ is the set of input points, $K(\bm{X},\bm{X})$ is the covariance matrix with components $k(x_i,x_j)$, $\bm{y}$ is the vector of data, $C$ is the covariance matrix of the data and $n$ is the number of data. After optimizing for $\sigma_f$ and $l$, the reconstructed function $\bm{f^*}$ at chosen points $\bm{X^*}$ can be given by
\be
<\bm{f^*}>=\bm{\mu^*}+K(\bm{X^*},\bm{X})[K(\bm{X},\bm{X})+C]^{-1}(\bm{y}-\bm{\mu})
\label{frec}
\ee
with
\be
\mathrm{cov}(\bm{f^*})=K(\bm{X^*},\bm{X^*})-K(\bm{X^*},\bm{X})[K(\bm{X},\bm{X})+C]^{-1}K(\bm{X},\bm{X^*})
\label{covfrec}
\ee

As we shall see, in order to reconstruct $q(z)$ from $H(z)$ and SNe Ia luminosity distances, we need to reconstruct also the function derivatives, so we will briefly explain how GP method furnishes this reconstruction.

\subsection{Reconstructing the derivative of a function}
The derivative of a GP is also a GP, thus one can obtain the covariance between the function and its derivative by differentiating the covariance function:
\be
\mathrm{cov}\left(f_i,\frac{\partial f_j}{\partial x_j}\right)=\frac{\partial k(x_i,x_j)}{\partial x_j}
\ee
and
\be
\mathrm{cov}\left(\frac{\partial f_i}{\partial x_i},\frac{\partial f_j}{\partial x_j}\right)=\frac{\partial^2k(x_i,x_j)}{\partial x_i\partial x_j}
\ee
Thus, we have
\be
f'(x)\sim\mathcal{GP}\left(\mu'(x),\frac{\partial^2k(x,\tilde{x})}{\partial x\partial \tilde{x}}\right)
\ee
and similarly for higher derivatives. With this GPs, the derivatives can thus be reconstructed in a similar way to Eqs. (\ref{frec}), (\ref{covfrec}). Further details can be found at \cite{seikel2012}.

\subsection{Combining \texorpdfstring{$f(x)$}{fx} and its derivatives}
When one is interested not only on reconstructing $f(x)$ and its derivatives, but also a function of them, like $g(f(x),f'(x),...)$ as in our case here, one needs to know the covariances between $f^{*}=f(x^{*})$, ${f^*}'=f'(x^{*})$ etc. at each point where $g$ is to be reconstructed. These covariances are given by:
\be
\mathrm{cov}(f^{*(i)},f^{*(j)})=k^{(i,j)}(x^*,x^*)-K^{(i)}(x^*,\bm{X})[K(\bm{X},\bm{X})+C]^{-1}K^{(j)}(\bm{X},x^*)
\label{covff1}
\ee
where $f^{*(i)}$ is the $i$th derivative of $f^*$ and $k^{(i,j)}(x^*,x^*)$ means that $k(x^*,x^*)$ is derived $i$ times with respect to the first argument and $j$ times with respect to the second argument.

Here, we use the Python package GaPP\footnote{Available at \url{http://www.acgc.uct.ac.za/~seikel/GAPP/index.html}. See \cite{seikel2012} for more details.} to reconstruct $q(z)$ from $H(z)$ and SNe Ia $D_L(z)$ data in order to obtain the transition redshift $z_t$ from the condition $q(z_t)=0$.
To reconstruct $q(z)$, we need up to first derivative $H'(z)$ and second derivative $D_L''(z)$, as we shall see in next section (Eqs. (\ref{qzH}) and (\ref{qzDL})). That is why we choose to work with the Gaussian, Mat\'ern(5/2) and Mat\'ern(7/2) kernels, which yield these derivatives, instead of Mat\'ern(3/2) kernel, which reconstructs only up to first derivative.

\section{\label{basic}Cosmological equations}
Our method consists basically on reconstructing the deceleration parameter, $q(z)$, from $H(z)$ data and from SNe Ia apparent magnitudes, $m_B(z)$, and then finding the transition redshift from the condition $q(z_t)=0$.

In order to perform this reconstruction, we must know how to obtain $q(z)$ from $H(z)$ and $m_B(z)$. Let us do it in the following.

The definition of the deceleration parameter $q(z)$ and its relation to $H(z)$ is given by: 
\begin{equation}
 q(z)\equiv-\frac{\ddot{a}}{aH^2}=\frac{1+z}{H}\frac{dH}{dz}-1\,, \label{qzH}
\end{equation}

As one may see, the deceleration parameter can be obtained from $H(z)$ and from its first derivative, $H'(z)$. The Gaussian Process, as performed by GaPP, furnish not only the $H(z)$ reconstruction from data, but also its derivatives up to fourth order and their estimated uncertainties and covariances.

So, it just remains to estimate the $q(z)$ uncertainty, $\sigma_q$, by error propagating Eq. (\ref{qzH}). We obtain:
\be
\left(\frac{\sigma_q}{1+q}\right)^2=\left(\frac{\sigma_{H '}}{H'}\right)^2+\left(\frac{\sigma_H}{H}\right)^2-\frac{2\sigma_{HH'}}{HH'}
\ee

In order to obtain $q(z)$ from SNe Ia apparent magnitudes, we use the relation between magnitude and distance luminosity:
\be
m=M+5\log d_L-5
\ee
where $m$ is apparent magnitude, $M$ is absolute magnitude and $d_L$ is distance luminosity in parsecs. 
By assuming a spatially flat Friedmann-Robertson-Walker cosmology, as preferred by inflation \cite{inflation} and indicated by current Cosmic Microwave Background observations \cite{Planck18}, the luminosity distance is given by:
\begin{equation}
\label{dl}
d_L(z)=(1+z)d_C(z),
\end{equation}
where $d_C$ is the comoving distance:
\begin{equation}
\label{dc}
d_C(z) = c\int_0^z \frac{dz}{H(z)},
\end{equation}
with $c$ being the speed of light $H(z)$ the Hubble parameter. For mathematical convenience, we choose to work with dimensionless quantities. Then, we define the dimensionless distances, $D_C\equiv\frac{d_C}{d_H}$, $D_L\equiv\frac{d_L}{d_H}$, $d_H\equiv c/H_0$ and the dimensionless Hubble parameter, $E(z)\equiv\frac{H(z)}{H_0}$. Thus, we have:
\begin{equation}
D_L(z)=(1+z)D_C(z),
\label{Dl}
\end{equation}
and
\begin{equation}
\label{Dc}
D_C(z) = \int_0^z \frac{dz}{E(z)},
\end{equation}
In terms of dimensionless distances, the apparent magnitude can be written:
\be
m=M+5\log\frac{c}{H_0}-5+5\log D_L=M_*+5\log D_L
\label{mDL}
\ee
where we have defined $M_*\equiv M+5\log\frac{c}{H_0}-5$. From (\ref{Dc}), it follows
\begin{equation}
 E(z)=\frac{1}{D_C'(z)},
 \label{EzDc}
\end{equation}
where a prime denotes derivative with respect to $z$. Eq. (\ref{qzH}) can also be written:
\be
q(z)=(1+z)\frac{E'}{E}-1=-(1+z)\frac{D_C''}{D_C'}-1
\label{qzDc}
\ee
With Eqs. (\ref{Dc}), (\ref{mDL}) and (\ref{qzDc}) we may find, after a straightforward but a bit tedious calculation:
\be
q(z)=\frac{2\alpha m'(1+z)-\alpha^2m'^2(1+z)^2-\alpha m''(1+z)^2-2}{\alpha m'(1+z)-1}-1
\ee
or:
\be
q(z)=\frac{1+\alpha m''(1+z)^2}{1-\alpha m'(1+z)}-\alpha m'(1+z)
\label{qzmz}
\ee
where $\alpha\equiv\frac{\ln10}{5}$. It is an interesting result, as it does not depend on $H_0$, neither on the magnitude normalization $M_*$.

However, we have found that the reconstruction of $m(z)$ does not yield reliable results. As explained in \cite{seikel2012}, given the same amount of data, functions that change very rapidly are more difficult to reconstruct than smooth functions. It happens that $m(z)$ is not an smooth function of the redshift. In fact, for any cosmological model, $D_L=z$ at low redshift, thus $m\sim M_*+5\log z$, so $m\rightarrow-\infty$ for $z\rightarrow0$.

So, we choose to work with $D_L(z)$, which is expected to be an smooth function of redshift, at least at low redshift, where most of SNe Ia data are located. From (\ref{mDL}), we can see that the luminosity distance can be given by:
\be
D_L=10^\frac{m-M_*}{5}
\label{DLm}
\ee
Thus, we can estimate the $D_L$ covariance matrix from the $m$ covariance matrix as:
\be
\Sigma_{d_L}=J\cdot\Sigma_m\cdot J^T
\ee
where $J$ is the Jacobian matrix of the variable change (\ref{DLm}). As $D_{Li}=D_L(m_i)$, the Jacobian is a diagonal matrix given by
\be
J=\mathrm{diag}(\sigma_{D_L}^2)=\mathrm{diag}\left[\left(\frac{dD_L}{dm}\right)^2\sigma_m^2\right]=\alpha^2\mathrm{diag}(D_{Li}^2\sigma_{mi}^2)
\ee
From (\ref{mDL}) and (\ref{qzmz}), we can find $q(z)$ as a function of $D_L(z)$ and its derivatives:
\be
q(z)=\frac{(1+z)^2D_L''}{D_L-(1+z)D_L'}+1
\label{qzDL}
\ee
As $D_L=10^\frac{m}{5}\times10^\frac{-M_*}{5}$, we can see that the $q(z)$ reconstruction will not depend on the scale $M_*$, as expected.

The $q(z)$ uncertainty can be estimated from the error propagation:
\bea
\left(\frac{D_L''}{q-1}\right)^2\sigma_q^2&=&\frac{(q-1)^2}{(1+z)^4}\sigma_{D_L}^2+\frac{(q-1)^2}{(1+z)^2}\sigma_{D_L'}^2+\sigma_{D_L''}^2-\nonumber\\
&-&\frac{2(q-1)^2}{(1+z)^3}\sigma_{D_LD_L'}-2\frac{(q-1)}{(1+z)^2}\sigma_{D_LD_L''}+2\frac{(q-1)}{(1+z)}\sigma_{D_L'D_L''}
\label{sigqz}
\eea

\section{\label{samples}Samples}
\subsection{\texorpdfstring{$H(z)$}{Hz}}
In order to reconstruct $q(z)$ from $H(z)$ data, we have considered the measurement of the Hubble parameter $H(z)$ in different redshifts. These kind of observational data are quite reliable because in general such observational data are independent of the background cosmological model, just relying on astrophysical assumptions. We have used the currently most complete compilation of $H(z)$ data, with 51 measurements \cite{MaganaEtAl18}.

Hubble parameter data as function of redshift yields one of the most straightforward cosmological tests because it is inferred from astrophysical observations alone, not depending on any background cosmological models.

At the present time, the most important methods for obtaining $H(z)$ data are\footnote{See \cite{limajesus2014} for a review.} (i) through ``cosmic chronometers", for example, the differential age of galaxies (DAG) \cite{Simon05,Stern10,Moresco12,Zhang12,Moresco15,MorescoEtAl16}, (ii) measurements of peaks of acoustic oscillations of baryons (BAO) \cite{Gazta09,Blake12,Busca12,AndersonEtAl13,Font-Ribera13,Delubac14} and (iii) through correlation function of luminous red galaxies (LRG) \cite{Chuang13,Oka14}.

Among these methods for estimating $H(z)$, the 51 data compilation as grouped by \cite{MaganaEtAl18}, consists of 20 clustering (BAO+LRG) and 31 differential age $H(z)$ data.

Differently from \cite{MaganaEtAl18}, we choose not to use $H_0$ in our main results here, due to the current tension among $H_0$ values estimated from different observations \citep{BernalEtAl16,Planck18,RiessEtAl19}.


\subsection{SNe Ia}
We choose to work with one of the largest SNe Ia sample to date, namely, the Pantheon sample \cite{pantheon}. This sample consists of 279 SNe Ia from Pan-STARRS1 (PS1) Medium Deep Survey ($0.03<z<0.68$), combined with distance estimates of SNe Ia from Sloan Digital Sky Survey (SDSS), SNLS and various low-$z$ and Hubble Space Telescope samples to form the largest combined sample of SNe Ia, consisting of a total of 1048 SNe Ia in the range of $0.01<z<2.3$. We take into account all the statistical and systematical SNe Ia uncertainties, as described by their full covariance matrix.

\section{\label{analysis}Analyses and Results}
\subsection{\texorpdfstring{$z_t$}{zt} from \texorpdfstring{$H(z)$}{Hz} reconstruction}
\begin{figure}
    \centering
    \includegraphics[width=.6\linewidth]{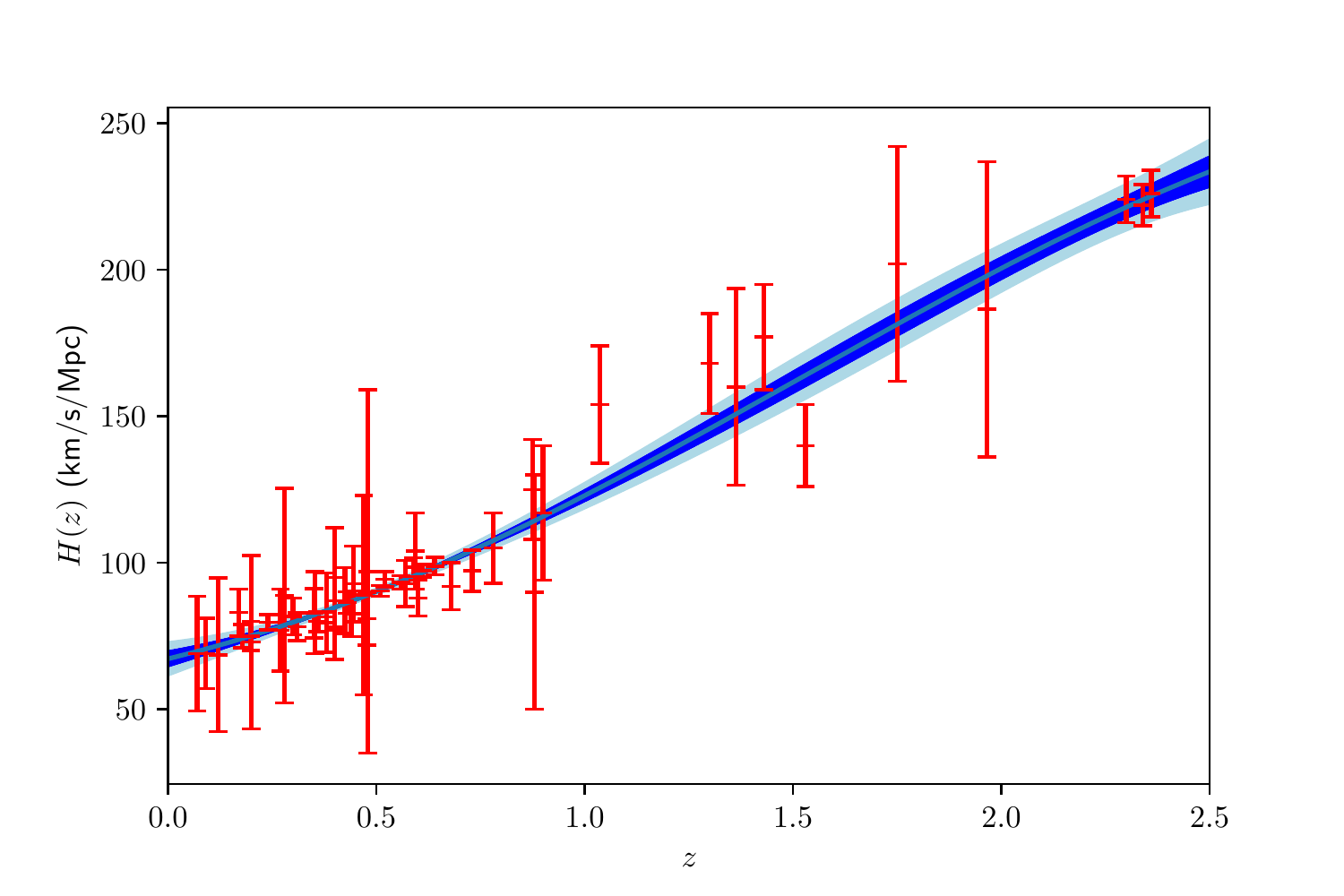}
    \includegraphics[width=.49\linewidth]{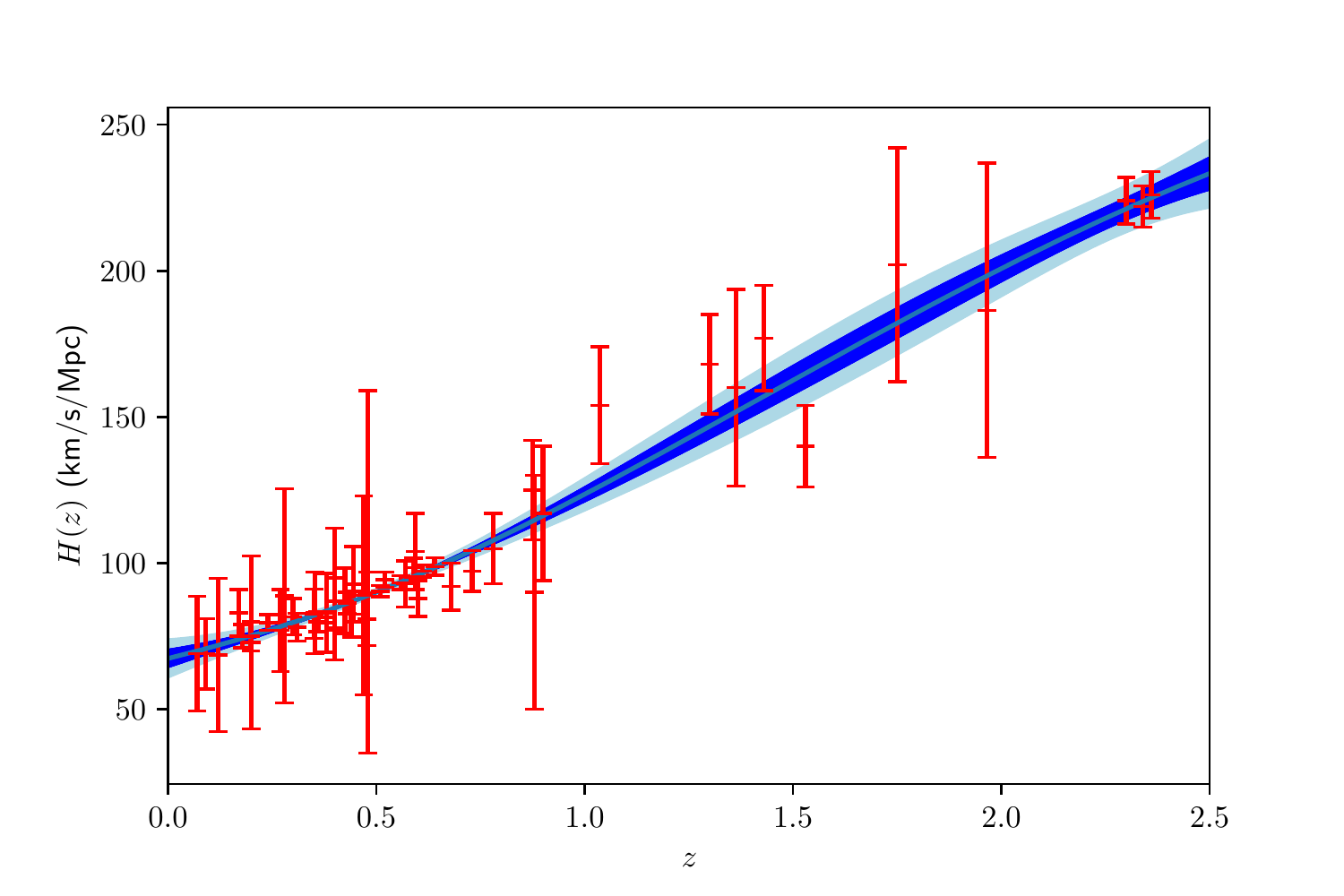}
    \includegraphics[width=.49\linewidth]{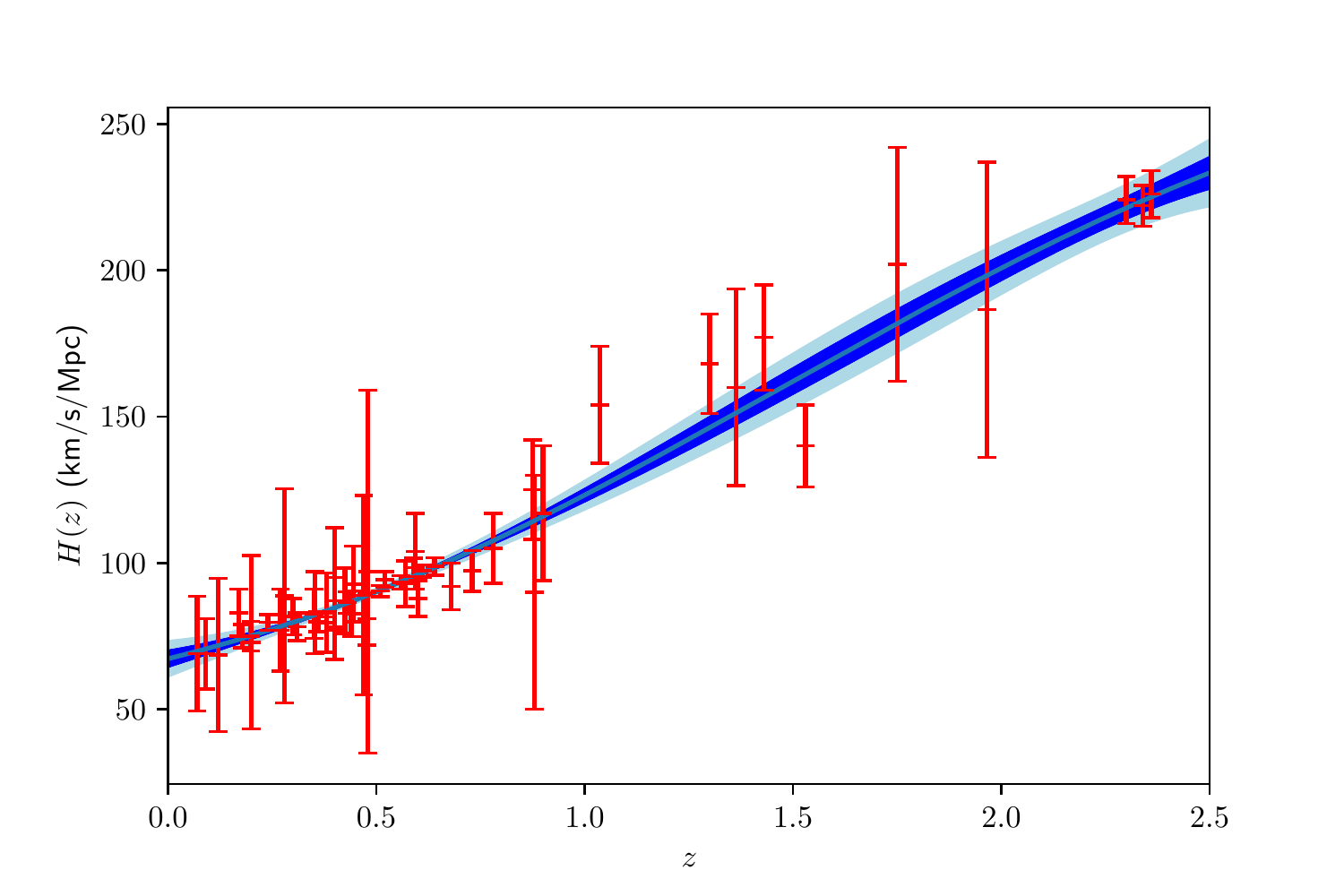}
    \caption{$H(z)$ reconstruction from covariance functions: Squared Exponential (top), Mat\'ern(5/2) (bottom-left) and Mat\'ern(7/2) (bottom-right), showing 68.3\% and 95.4\% c.l., together with 51 $H(z)$ data.}
    \label{hzrechz}
\end{figure}

As explained above, we can reconstruct $H(z)$ directly from data, by using the Gaussian Processes method. The result of this reconstruction for the 51 available $H(z)$ data can be seen on Fig. \ref{hzrechz}, within 1 and 2 $\sigma$ (68.3\% and 95.4\% c.l.) confidence intervals.

\begin{figure}
    \includegraphics[width=.6\linewidth]{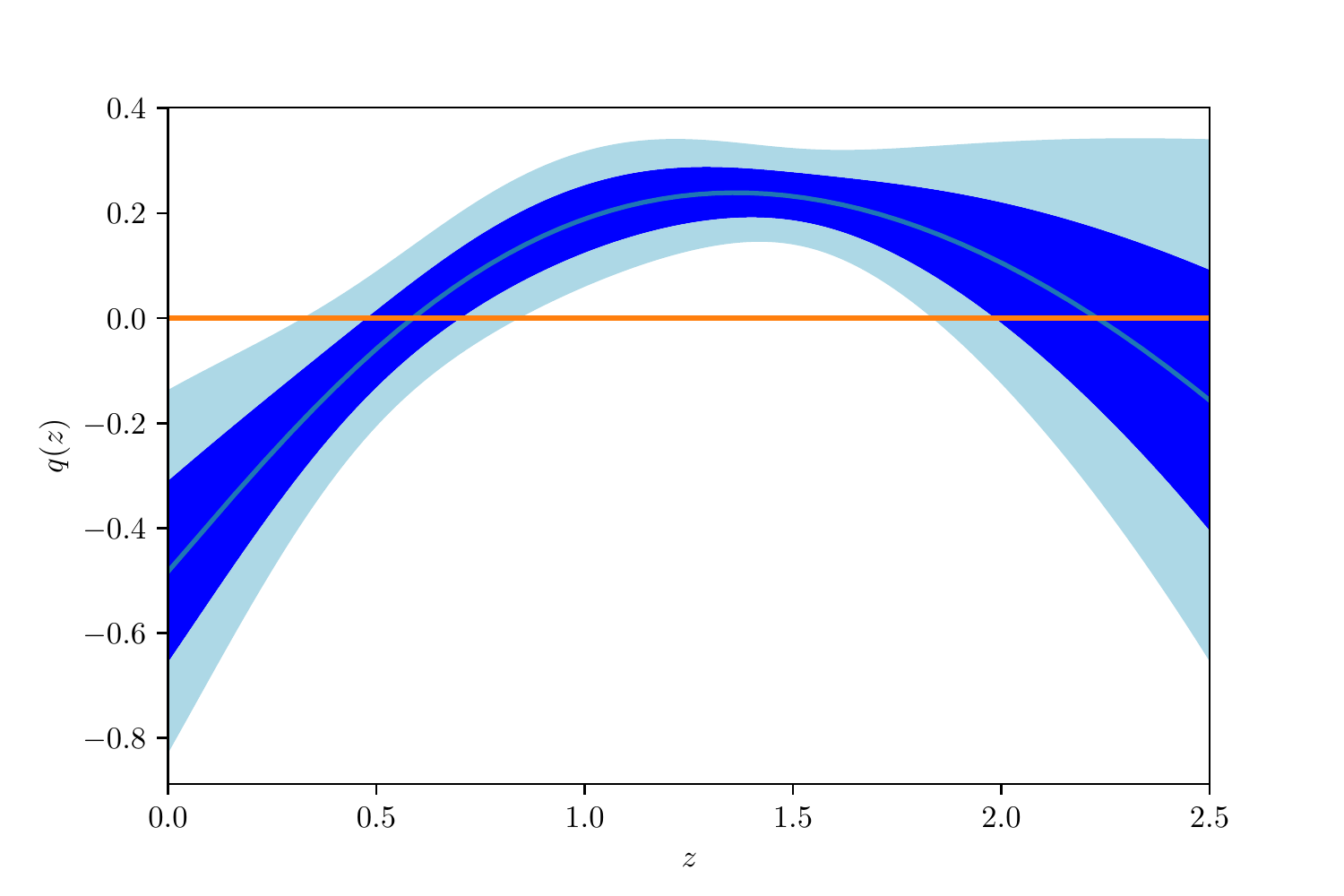}
    \includegraphics[width=.49\linewidth]{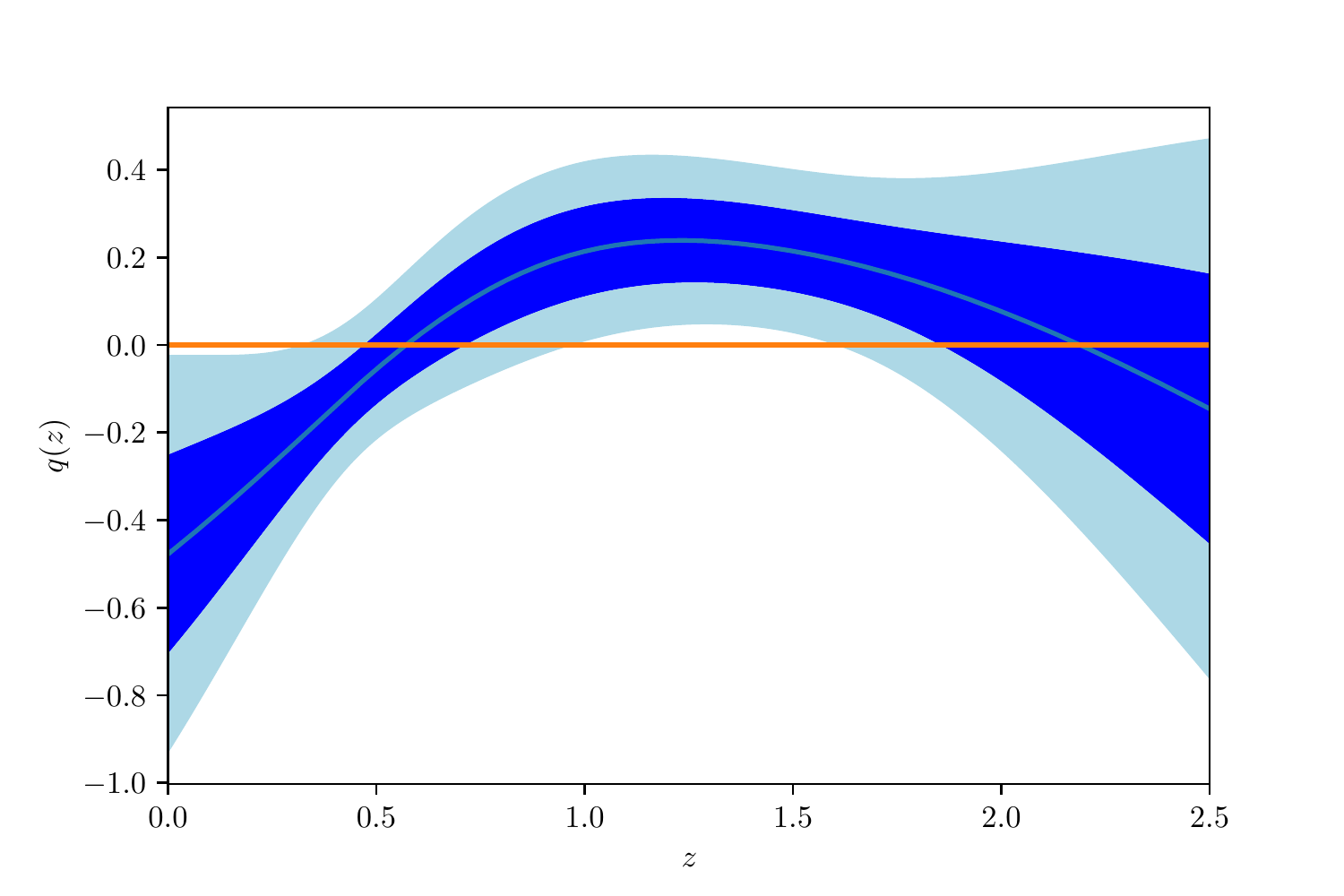}
    \includegraphics[width=.49\linewidth]{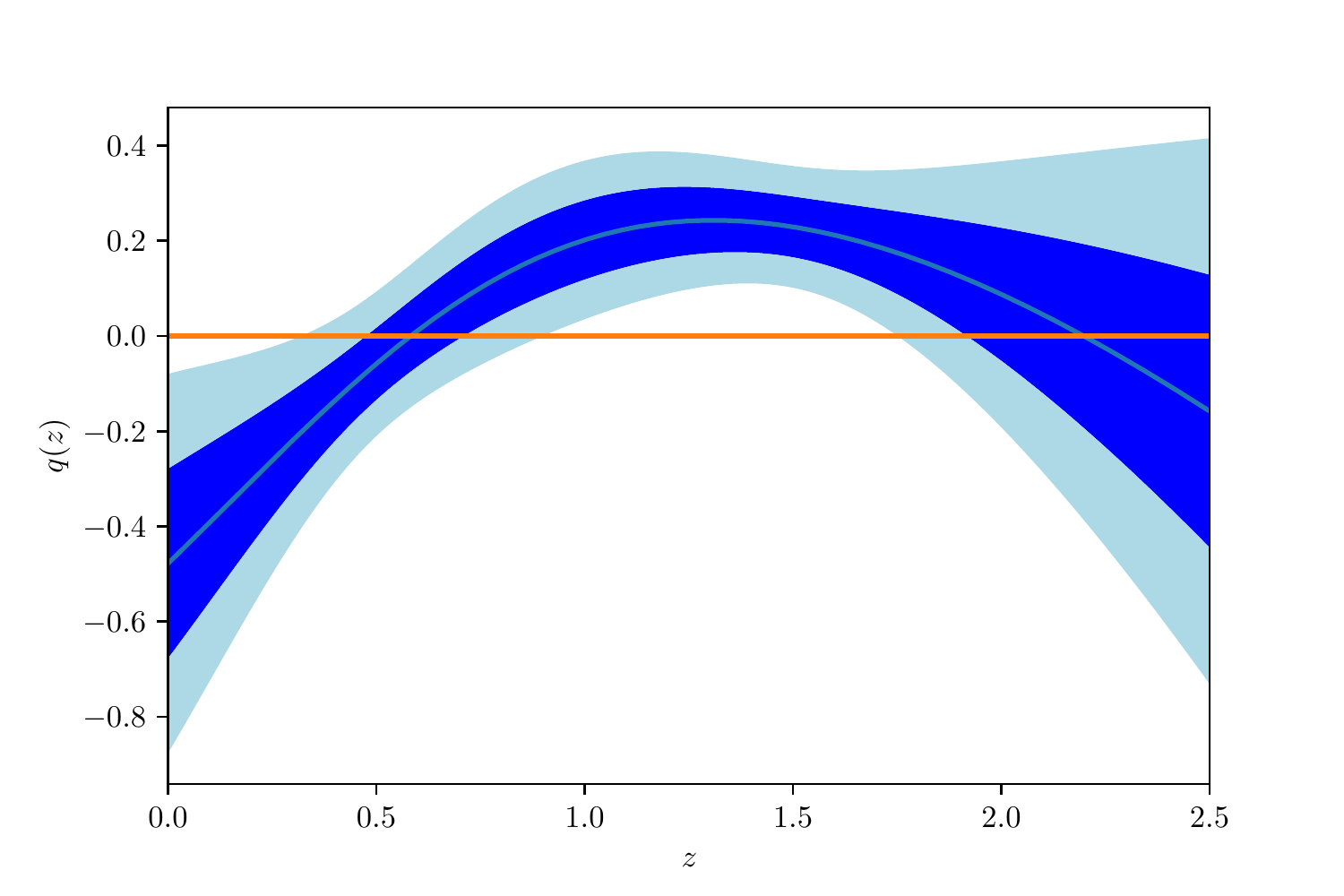}
    \caption{$q(z)$ reconstruction from 51 $H(z)$ data, using covariance functions: Squared Exponential (top), Mat\'ern(5/2) (bottom-left) and Mat\'ern(7/2) (bottom-right), showing 68.3\% and 95.4\% c.l.}
    \label{qzrechz}
\end{figure}

From the reconstruction of $H(z)$ and $H'(z)$, we have reconstructed the deceleration parameter, $q(z)$, according to Eq. (\ref{qzH}). The result of this reconstruction can be seen on Fig. \ref{qzrechz}.

The transition redshift can be found from this reconstruction with the condition $q(z_t)=0$, thus corresponding to the intersection with the orange line in Fig. \ref{qzrechz}.

In principle, we should determine the $q(z)$ uncertainties by Monte Carlo sampling a multivariate normal distribution with the covariance matrix given by (\ref{covff1}), but we found negligible difference with the error propagation (\ref{sigqz}), over the full range $0<z<2.5$, so we used the error propagation, due to simplicity.

\subsection{\texorpdfstring{$z_t$}{zt} from \texorpdfstring{$D_L(z)$}{DL} reconstruction}
\begin{figure}[ht]
 \includegraphics[width=.6\linewidth]{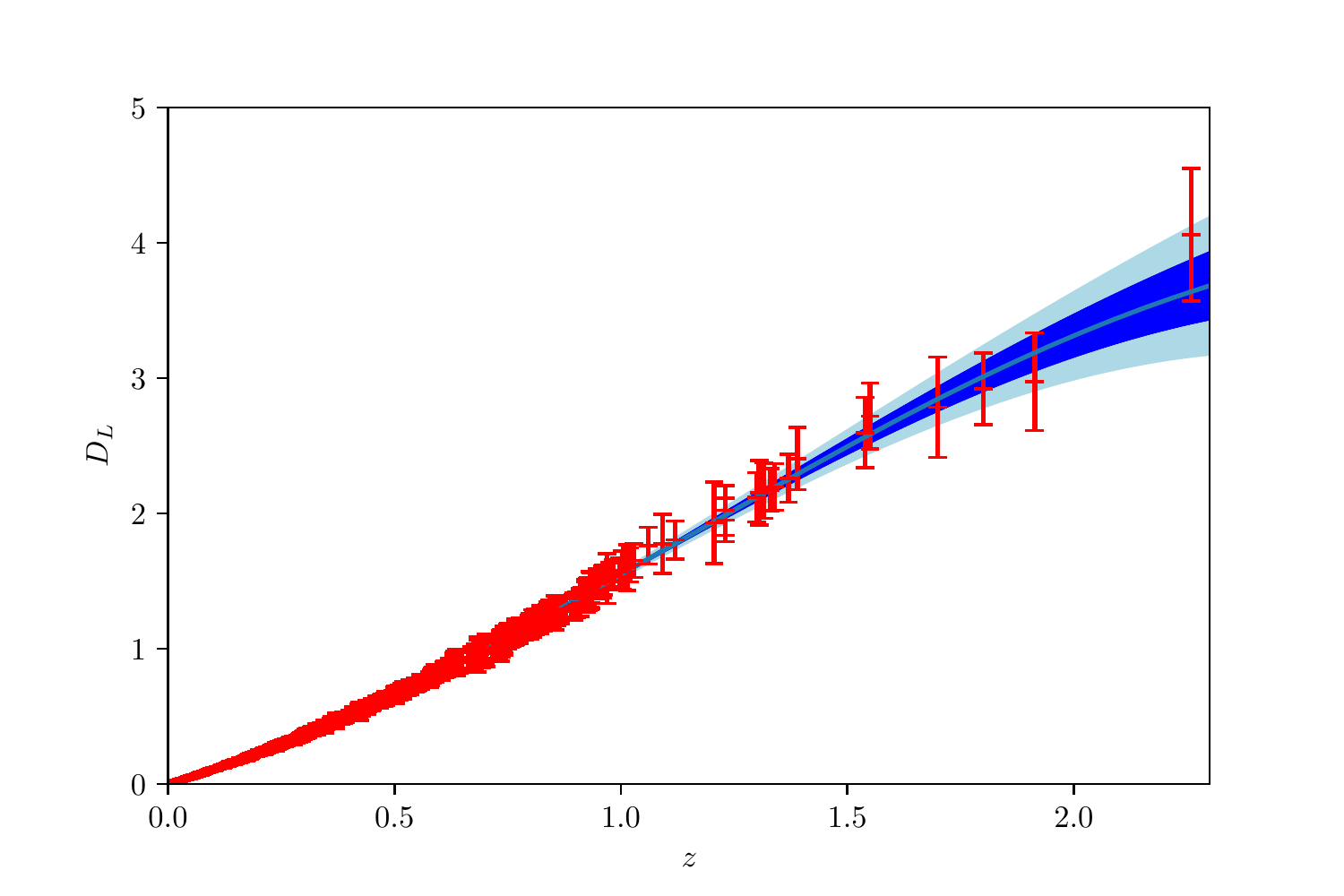}
 \includegraphics[width=.49\linewidth]{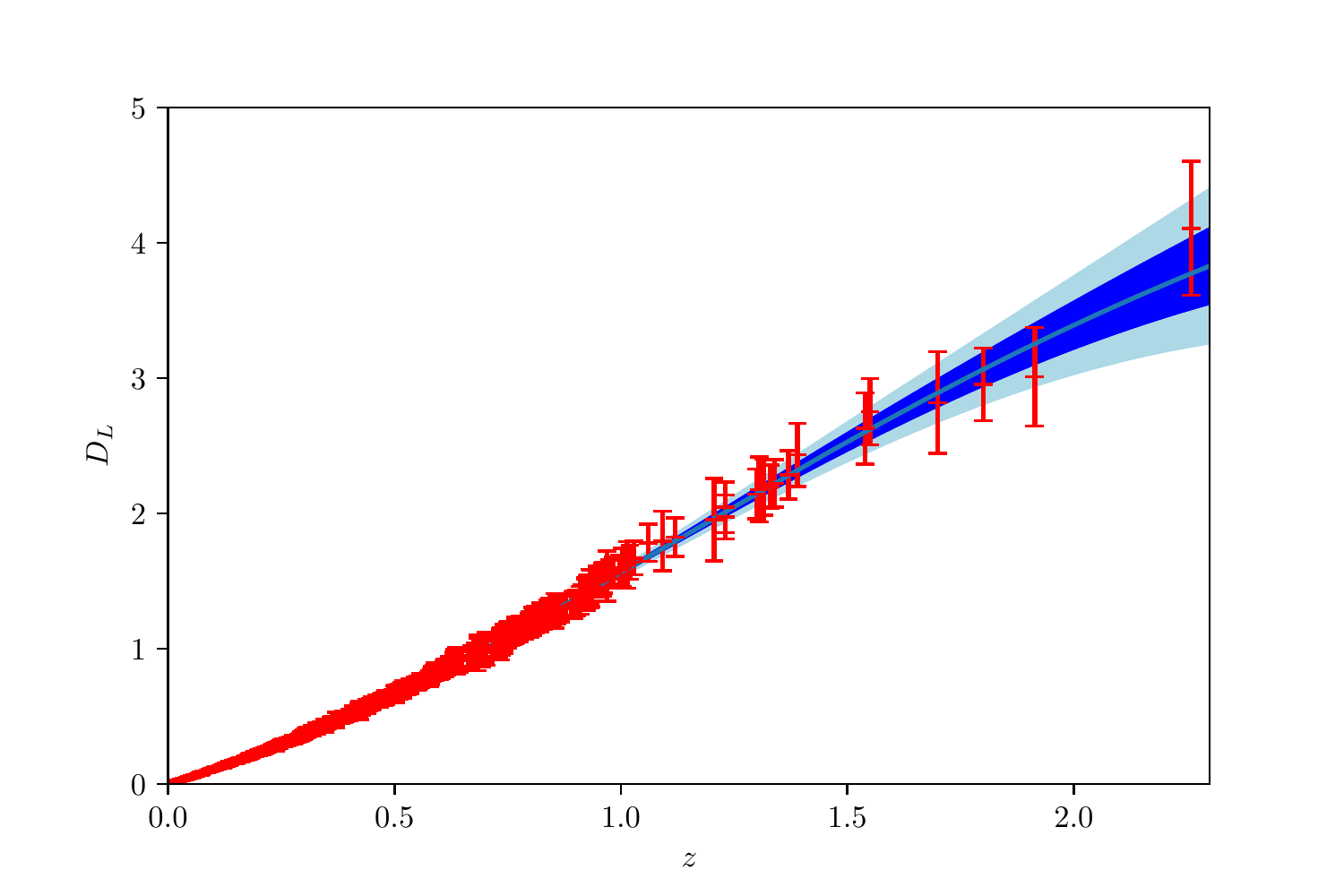}
 \includegraphics[width=.49\linewidth]{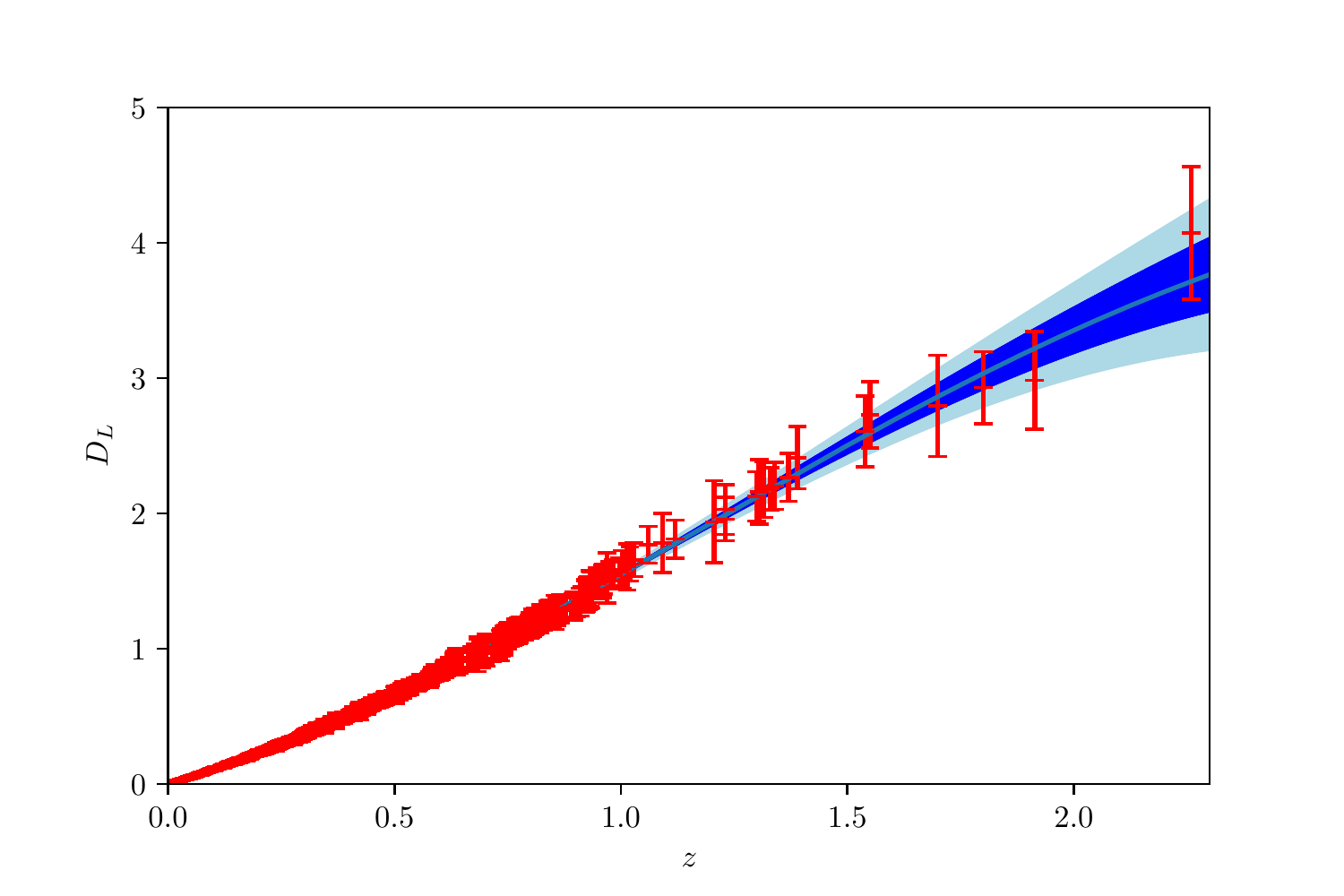}
 \caption{\label{DLrecPanth} Reconstruction of $D_L(z)$ from Pantheon SNe Ia data. Squared Exponential (top), Mat\'ern(5/2) (bottom-left) and Mat\'ern(7/2) (bottom-right).}
\end{figure}

The result of the reconstruction of the luminosity distance for the 1048 Pantheon SNe Ia data can be seen on Fig. \ref{DLrecPanth}, within 1 and 2 $\sigma$ (68.3\% and 95.4\% c.l.) confidence intervals.

\newpage

\begin{figure}[ht]
 \includegraphics[width=.6\linewidth]{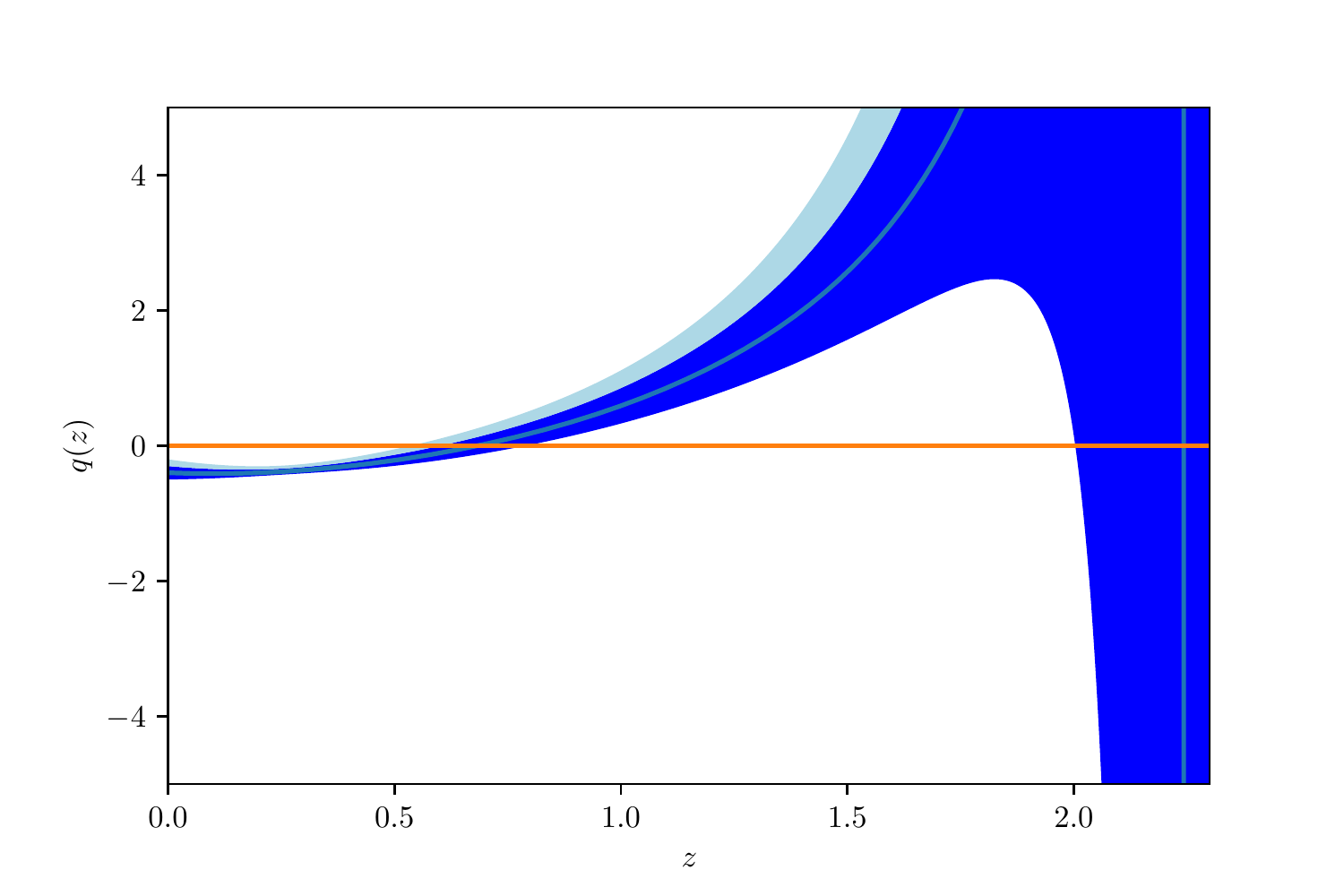}
 \includegraphics[width=.49\linewidth]{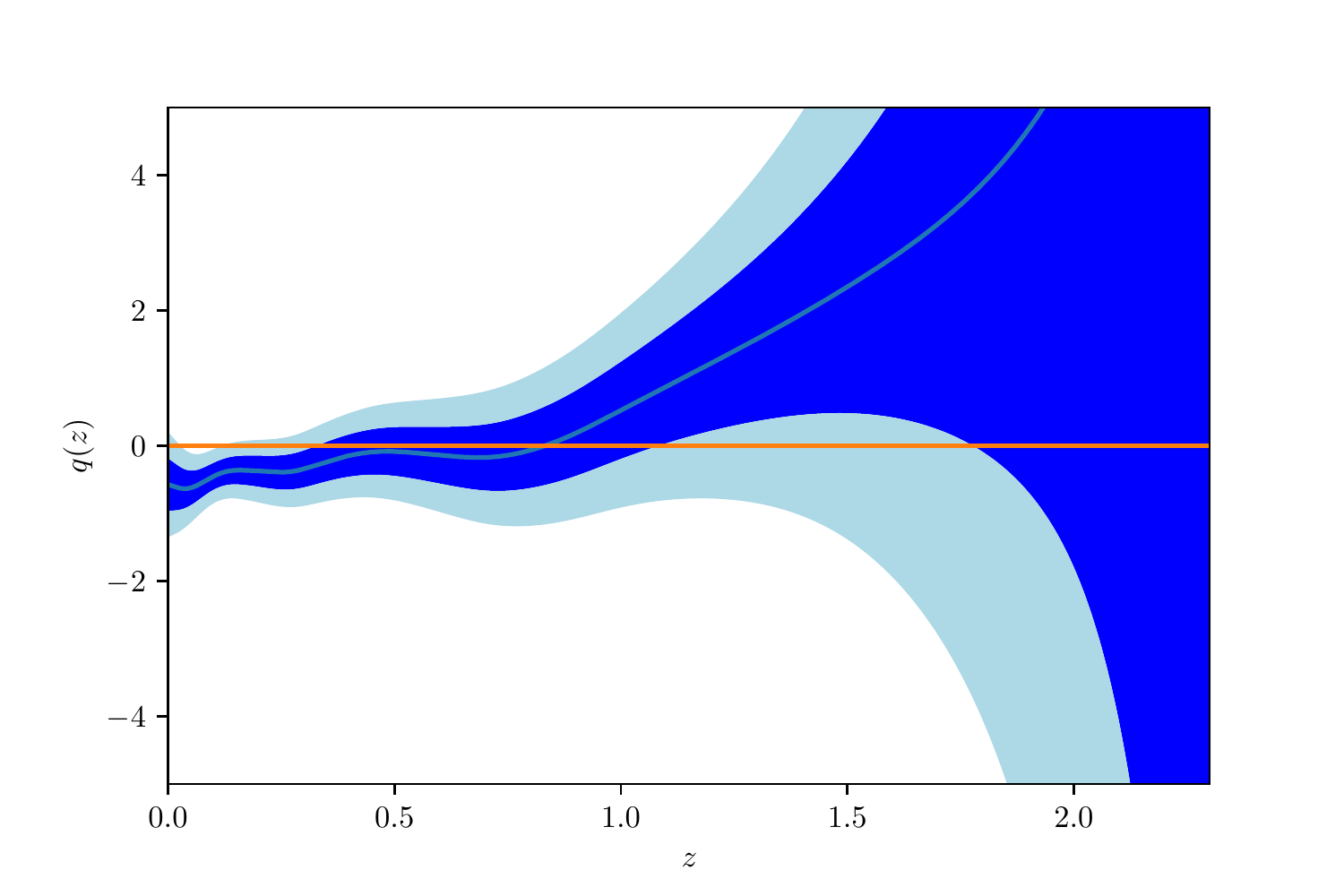}
 \includegraphics[width=.49\linewidth]{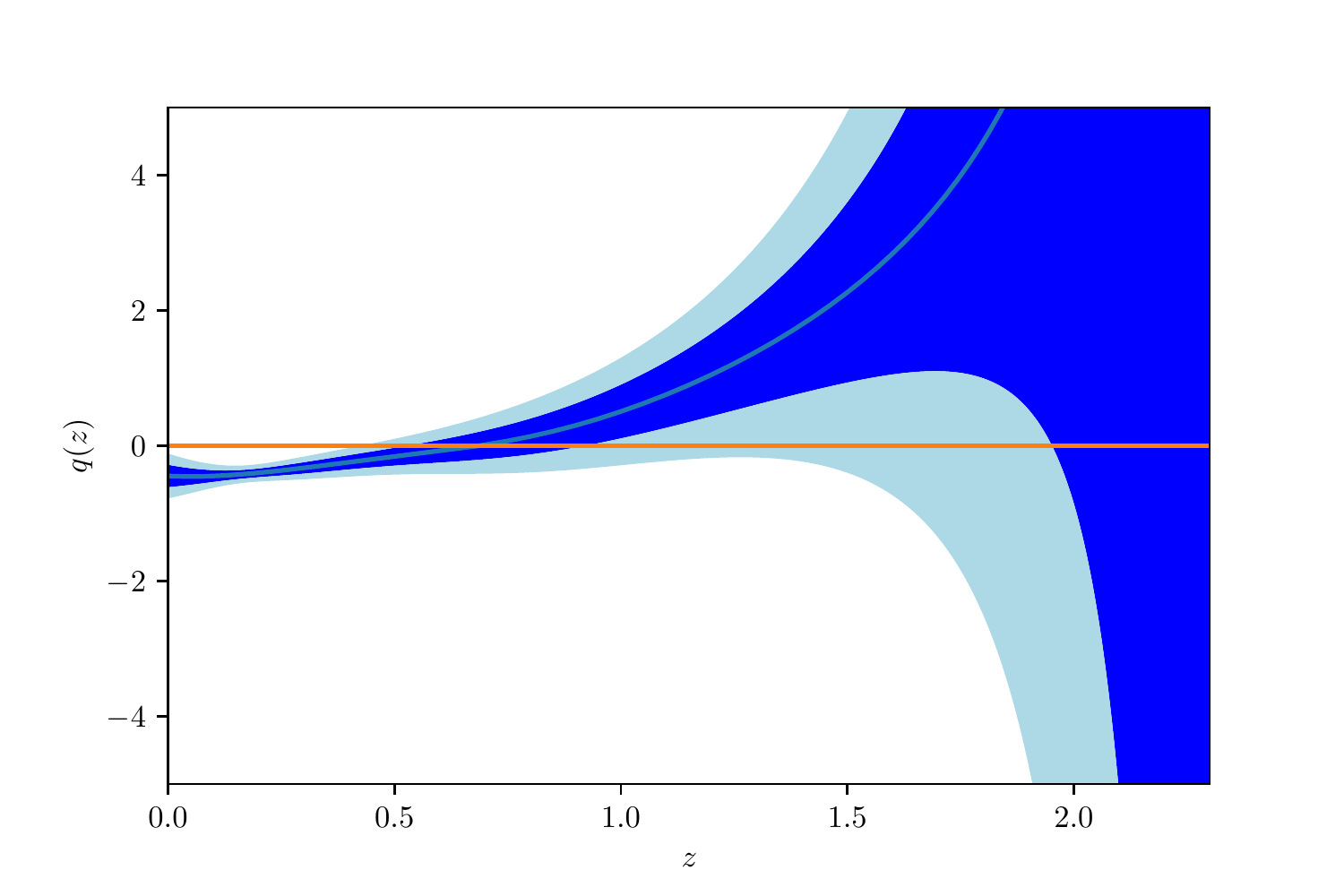}
 \caption{\label{qzrecPanth} Reconstruction of $q(z)$ from Pantheon SNe Ia data. Squared Exponential (top), Mat\'ern(5/2) (bottom-left) and Mat\'ern(7/2) (bottom-right).}
\end{figure}

From the reconstruction of $D_L(z)$, $D_L'(z)$ and $D_L''(z)$, we have reconstructed the deceleration parameter, $q(z)$, according to Eq. (\ref{qzDL}). The result of this reconstruction can be seen on Fig. \ref{qzrecPanth}.

As mentioned above, according to \cite{seikel2012}, the uncertainties on $q(z)$ should be estimated by Monte Carlo sampling the multivariate normal distribution of $D_L(z)$, $D_L'(z)$ and $D_L''(z)$, mainly when one has large uncertainties. We have tested it, but the difference with the error propagation (\ref{sigqz}) is negligible for $z\lesssim1$, thus, not influencing the $z_t$ determination. It is expected, as the most of SNe Ia data are available at lower redshifts, yielding lower $q(z)$ uncertainties. We then choose to work with the error propagation (\ref{sigqz}), for simplicity.

\newpage

Table \ref{tab1} shows the full numerical results from our statistical analysis. Here, it may be clearer how the SNe Ia parameters varies little for each kernel. One can also see that, for each kernel, the constraints over $z_t$ from $H(z)$ and $D_L(z)$ are compatible at 68.3\% c.l. The poorest constraints come from $D_L(z)$, Mat\'ern(5/2) kernel and the best constraints come from $D_L(z)$, Squared Exponential kernel.

\begin{table}[ht]
\begin{tabular}{l|c|c|c|}
\cline{2-3}
&\multicolumn{2}{|c|}{$z_t$} \\
\hline
\multicolumn{1}{|c|}{Method} & $H(z)$ & $D_L(z)$         \\
\hline
\multicolumn{1}{|c|}{Sq. Exp.} & $0.59^{+0.12+0.26}_{-0.11-0.26}$  & $0.683^{+0.11\,\,\,+0.26}_{-0.082-0.15}$\\
\multicolumn{1}{|c|}{Mat\'ern(5/2)} & $0.57^{+0.14+0.40}_{-0.10-0.25}$  & $0.83^{+0.25}_{-0.50}$\\
\multicolumn{1}{|c|}{Mat\'ern(7/2)} & $0.58^{+0.13+0.32}_{-0.11-0.26}$  & $0.69^{+0.23}_{-0.16}$\\
\hline
\end{tabular}
\caption{Constraints from $D_L(z)$ and $H(z)$ data. The 1 $\sigma$ and 2 $\sigma$ c.l. correspond to the minimal 68.3\% and 95.4\% confidence intervals.}
\label{tab1}
\end{table}

By using 30 $H(z)$ data, plus $H_0$ from Riess et al. (2011) \cite{Riess11}, Moresco et al. \cite{moresco2016} found $z_t=0.64^{+0.1}_{-0.06}$ for $\Lambda$CDM and $z_t=0.4\pm0.1$ for their model independent approach. These results are both compatible with our model-independent result. The kinematic parametrizations of \cite{jrs2018} yielded $z_t=0.806\pm 0.094$, $0.870\pm 0.063$ and $0.973\pm 0.058$ at 1$\sigma$ c.l. for $D_C(z)$, $H(z)$ and $q(z)$ parametrizations, respectively. It is a result compatible with our $D_L(z)$ Mat\'ern(5/2) and Mat\'ern(7/2) reconstructions. They are incompatible with all our $H(z)$ reconstructions at 1$\sigma$ c.l., although compatible at 2$\sigma$ c.l. Our $D_L(z)$ Sq. Exp. kernel result is compatible only with the $D_C(z)$ kinematic parametrization.
Using GP with 36 $H(z)$ data, Yu et al. \cite{YuEtAl18} find $0.33<z_t<1$ at 1$\sigma$ c.l., compatible with our results.


\section{\label{conclusion}Conclusion}

The transition from decelerated to the current accelerated phase of expansion of the universe is an important question in modern cosmology. It is well known that the transition redshift  $ z_t $ is strongly model dependent, however some recent methods allow estimating cosmological parameters in model independent approaches. The so called Gaussian Process method furnish an interesting way to reconstruct some cosmological functions based just on observational data, with no need of a priori model. Given the increasing amount of recent cosmological data, it is a powerful tool to test theoretical models.

In the present work we have found constraints over the transition redshift $ z_t $ from $H(z)$ and SNe Ia data, by using the non-parametric Gaussian Process method. It was done the reconstruction of $H(z)$ function and its derivatives up to third order together its uncertainties, and then the transition redshift was obtained from decelerator parameter $q(z)$. The same procedure was done with SNe Ia, where the module distance function was reconstructed from luminosity distance. Our SNe Ia result relies on the spatial flatness assumption.

The main results are summarized in Table I. The deceleration parameter reconstruction from $H(z)$ data yields $z_t=0.59^{+0.12}_{-0.11}$. The reconstruction from SNe Ia data assumes spatial flatness and yields $z_t=0.683^{+0.11}_{-0.082}$. For each kernel, namely Squared Exponential, Mat\'ern(5/2) and Mat\'ern(7/2), the constraints over $z_t$ from $H(z)$ and $D_L(z)$ are compatible at 1 $\sigma$ c.l. The best constraints come from $D_L(z)$ for Squared Exponential kernel while the poorest one come from $D_L(z)$ at Mat\'ern(5/2) kernel.

\begin{acknowledgments}
JFJ is supported by Funda\c{c}\~ao de Amparo \`a Pesquisa do Estado de S\~ao Paulo - FAPESP (Process number 2017/05859-0). RV is supported by  Funda\c{c}\~ao de Amparo \`a Pesquisa do Estado de S\~ao Paulo - FAPESP (thematic project process no. 2013/26258-2 and regular project process no. 2016/09831-0). SHP acknowledges financial support from  {Conselho Nacional de Desenvolvimento Cient\'ifico e Tecnol\'ogico} (CNPq)  (No. 303583/2018-5 and 400924/2016-1). This study was financed in part by the Coordena\c{c}\~ao de Aperfei\c{c}oamento de Pessoal de N\'ivel Superior - Brasil (CAPES) - Finance Code 001.
\end{acknowledgments}


\begin{thebibliography}{30}



\bibitem{SN1}
  A.~G.~Riess {\it et al.} [Supernova Search Team],
  Astron.\ J.\  {\bf 116} (1998) 1009
  [astro-ph/9805201].

\bibitem{SN2}
  S.~Perlmutter {\it et al.} [Supernova Cosmology Project Collaboration],
  Astrophys.\ J.\  {\bf 517} (1999) 565
  [astro-ph/9812133].

\bibitem{SN3}
  P.~Astier {\it et al.} [SNLS Collaboration],
  Astron.\ Astrophys.\  {\bf 447} (2006) 31
  [astro-ph/0510447].

\bibitem{SN4}
  A.~G.~Riess {\it et al.},
  Astrophys.\ J.\  {\bf 659} (2007) 98
  [astro-ph/0611572].

\bibitem{SN5}
  T.~M.~Davis {\it et al.},
  Astrophys.\ J.\  {\bf 666} (2007) 716
  [astro-ph/0701510].

\bibitem{union}
  M.~Kowalski {\it et al.} [Supernova Cosmology Project Collaboration],
  Astrophys.\ J.\  {\bf 686} (2008) 749
  [arXiv:0804.4142 [astro-ph]].

\bibitem{union2}
  R.~Amanullah {\it et al.},
  Astrophys.\ J.\  {\bf 716} (2010) 712
  [arXiv:1004.1711 [astro-ph.CO]].

\bibitem{union21}
  N.~Suzuki {\it et al.},
  Astrophys.\ J.\  {\bf 746} (2012) 85
  [arXiv:1105.3470 [astro-ph.CO]].

\bibitem{WMAP1}
  E.~Komatsu {\it et al.} [WMAP Collaboration],
  Astrophys.\ J.\ Suppl.\  {\bf 192} (2011) 18
  [arXiv:1001.4538 [astro-ph.CO]].

\bibitem{WMAP2}
  D.~Larson {\it et al.},
  Astrophys.\ J.\ Suppl.\  {\bf 192} (2011) 16
  [arXiv:1001.4635 [astro-ph.CO]].

\bibitem{planck}
  P.~A.~R.~Ade {\it et al.} [Planck Collaboration],
  Astron.\ Astrophys.\  {\bf 571} (2014) A16
  [arXiv:1303.5076 [astro-ph.CO]].
  
\bibitem{BAO1}
  D.~J.~Eisenstein {\it et al.} [SDSS Collaboration],
  Astrophys.\ J.\  {\bf 633} (2005) 560
  [astro-ph/0501171].

\bibitem{BAO2}
  W.~J.~Percival, S.~Cole, D.~J.~Eisenstein, R.~C.~Nichol, J.~A.~Peacock, A.~C.~Pope and A.~S.~Szalay,
  Mon.\ Not.\ Roy.\ Astron.\ Soc.\  {\bf 381} (2007) 1053
  [arXiv:0705.3323 [astro-ph]].

\bibitem{BAO3}
  D.~Schlegel {\it et al.} [with input from the SDSS-III Collaboration],
  arXiv:0902.4680 [astro-ph.CO].

\bibitem{BAO4}
  D.~J.~Eisenstein {\it et al.} [SDSS Collaboration],
  Astron.\ J.\  {\bf 142} (2011) 72
  [arXiv:1101.1529 [astro-ph.IM]].

\bibitem{BAO5}
  K.~S.~Dawson {\it et al.} [BOSS Collaboration],
  Astron.\ J.\  {\bf 145} (2013) 10
  [arXiv:1208.0022 [astro-ph.CO]].

  
\bibitem{Omer}
  O.~Farooq, D.~Mania and B.~Ratra,
  Astrophys.\ J.\  {\bf 764} (2013) 138
  [arXiv:1211.4253 [astro-ph.CO]].

\bibitem{Omer2}
  O.~Farooq and B.~Ratra,
  Astrophys.\ J.\  {\bf 766} (2013) L7
  [arXiv:1301.5243 [astro-ph.CO]].

\bibitem{Omer3}
  O.~Farooq, F.~R.~Madiyar, S.~Crandall and B.~Ratra,
  Astrophys.\ J.\  {\bf 835} (2017) no.1,  26
  [arXiv:1607.03537 [astro-ph.CO]].


\bibitem{sharov}
  G.~S.~Sharov and E.~G.~Vorontsova,
  JCAP {\bf 1410} (2014) 10,  057
  [arXiv:1407.5405 [gr-qc]].

\bibitem{kine3}
  C.~Shapiro and M.~S.~Turner,
  Astrophys.\ J.\  {\bf 649} (2006) 563
  [astro-ph/0512586].



\bibitem{cunha2008}
  J.~V.~Cunha and J.~A.~S.~Lima,
  Mon.\ Not.\ Roy.\ Astron.\ Soc.\  {\bf 390} (2008) 210
  [arXiv:0805.1261 [astro-ph]].

\bibitem{guimaraes2009}
  A.~C.~C.~Guimaraes, J.~V.~Cunha and J.~A.~S.~Lima,
  JCAP {\bf 0910} (2009) 010
  [arXiv:0904.3550 [astro-ph.CO]].

\bibitem{Rani2015}
  N.~Rani, D.~Jain, S.~Mahajan, A.~Mukherjee and N.~Pires,
  JCAP {\bf 1512} (2015) no.12,  045
  [arXiv:1503.08543 [gr-qc]].


\bibitem{limajesus2014}
  J.~A.~S.~Lima, J.~F.~Jesus, R.~C.~Santos and M.~S.~S.~Gill,
  arXiv:1205.4688 [astro-ph.CO].

\bibitem{moresco2016}
	M. Moresco {\it et al.},
	JCAP {\bf 1605} (2016) 05,  014
	[arXiv:1601.01701 [astro-ph.CO]].
	
	
\bibitem{jrs2018}J.F. Jesus, R.F.L. Holanda and S.H. Pereira, JCAP 05 (2018) 073.


\bibitem{Xu2009}
  L.~Xu, W.~Li and J.~Lu,
  JCAP {\bf 0907} (2009) 031
  [arXiv:0905.4552 [astro-ph.CO]].

\bibitem{JLA}
  M.~Betoule {\it et al.} [SDSS Collaboration],
  Astron.\ Astrophys.\  {\bf 568} (2014) A22
  [arXiv:1401.4064 [astro-ph.CO]].
  
\bibitem{seikel2012}  M. Seikel, C. Clarkson and M. Smith, 
JCAP {\bf 06} (2012) 036.

\bibitem{Rasmussen}C. Rasmussen and C. Williams, \emph{Gaussian Processes for Machine Learning}, MIT Press,
Cambridge U.S.A. (2006).

\bibitem{Hols1}T. Holsclaw et al., 
Phys. Rev. Lett. 105 (2010) 241302.

\bibitem{Hols2}T. Holsclaw et al., 
Phys. Rev. D 82 (2010) 103502.

\bibitem{Shafieloo}
A. Shafieloo, A. G. Kim and E. V. Linder,
Phys. Rev. D 85, (2012) 123530
 
\bibitem{Yin}Zhao-Yu Yin and H. Wei,
arXiv:1808.00377 [astro-ph.CO].

\bibitem{Yenn1}M. K. Yennapureddy and F. Melia, 
Eur. Phys. J. C (2018) 78: 258


\bibitem{Yenn2}M. K. Yennapureddy and F. Melia,
JCAP 11, 029 (2017).

\bibitem{Yenn3}F. Melia and M. K. Yennapureddy,
JCAP 02, 034 (2018).

\bibitem{melia2012}F. Melia and A. Shevchuk,
MNRAS 419 2579.

\bibitem{inflation}
  A.~H.~Guth,
  Phys.\ Rev.\ D {\bf 23} (1981) 347
   [Adv.\ Ser.\ Astrophys.\ Cosmol.\  {\bf 3} (1987) 139].

\bibitem{Planck18}
  N.~Aghanim {\it et al.} [Planck Collaboration],
  arXiv:1807.06209 [astro-ph.CO].

\bibitem{MaganaEtAl18}
  J.~Magana, M.~H.~Amante, M.~A.~Garcia-Aspeitia and V.~Motta,
  Mon.\ Not.\ Roy.\ Astron.\ Soc.\  {\bf 476} (2018) no.1,  1036
  [arXiv:1706.09848 [astro-ph.CO]].

\bibitem{Simon05}
 J. Simon, L. Verde and R. Jimenez, \emph{Constraints on the redshift dependence of the dark energy
potential}, \emph{Phys. Rev. {\bf D}} {\bf 71} (2005) 123001
[astro-ph/0412269].

\bibitem{Stern10}
D. Stern,  R. Jimenez,  L. Verde, M. Kamionkowski and S. A. Stanford,
\emph{Cosmic chronometers: constraining the equation of state of dark energy.
I: $H(z)$ measurements}, \emph{J. of Cosmology and Astropart. Phys.} {\bf 02}
(2010) 008 [arXiv:0907.3149].

\bibitem{Moresco12}
M. Moresco {\it et al.}, \emph{Improved constraints on the expansion rate of the
Universe up to $z$~1.1 from the spectroscopic evolution of cosmic
chronometers}, \emph{J. of Cosmology and Astropart. Phys.} {\bf 8} (2012) 006
[arXiv:1201.3609].

\bibitem{Zhang12} 
  C.~Zhang, H.~Zhang, S.~Yuan, T.~J.~Zhang and Y.~C.~Sun,
  Res.\ Astron.\ Astrophys.\  {\bf 14}, no. 10, 1221 (2014)
  [arXiv:1207.4541 [astro-ph.CO]].

\bibitem{Moresco15}
M. Moresco, \emph{Raising the bar: new constraints on the Hubble parameter with cosmic chronometers at z $\approx$ 2,},
\emph{Mon. Not. Roy. Astron. Soc.} {\bf 450} (2015) L16
[arXiv:1503.01116].

\bibitem{MorescoEtAl16}
  M.~Moresco {\it et al.},
  JCAP {\bf 1605} (2016) no.05,  014
  [arXiv:1601.01701 [astro-ph.CO]].
  
\bibitem{Gazta09}
E. Gazta\~naga,  A. Cabre, L. Hui, \emph{Clustering of Luminous Red Galaxies
IV: Baryon Acoustic Peak in the Line-of-Sight Direction
and a Direct Measurement of $H(z)$}, \emph{Mon. Not. Roy. Astron. Soc.} {\bf 399(3)} (2009) 1663 
[arXiv:0807.3551].

\bibitem{Blake12}
C. Blake {\it et al.}, \emph{The WiggleZ Dark Energy Survey: Joint measurements of
the expansion and growth history at
$z < 1$ }, \emph{Mon. Not. Roy. Astron. Soc.} {\bf 425(1)} (2012) 405 
[arXiv:1204.3674]. 

\bibitem{Busca12}
 N. G. Busca {\it et al.}, \emph{Baryon Acoustic Oscillations in the Ly$\alpha$ forest of BOSS quasars}, \emph{Astron. and Astrop.}
 {\bf 552} (2013) A96 [arXiv:1211.2616].

\bibitem{AndersonEtAl13} 
  L.~Anderson {\it et al.},
  Mon.\ Not.\ Roy.\ Astron.\ Soc.\  {\bf 439}, no. 1, 83 (2014)
  [arXiv:1303.4666 [astro-ph.CO]].

\bibitem{Font-Ribera13}
A. Font-Ribera {\it et al.}, \emph{Quasar-Lyman $\alpha$ Forest
Cross-Correlation from BOSS DR11: Baryon Acoustic Oscillations},
\emph{J. of Cosmology and Astroparticle Phys.} {\bf 05} (2014) 027
[arXiv:1311.1767]. 

\bibitem{Delubac14}
  T.~Delubac {\it et al.} [BOSS Collaboration],
  Astron.\ Astrophys.\  {\bf 574} (2015) A59
  [arXiv:1404.1801 [astro-ph.CO]].

\bibitem{Chuang13}
C.H. Chuang and Y. Wang, \emph{Modeling the Anisotropic Two-Point
Galaxy Correlation Function on Small Scales and Improved
Measurements of $H(z)$, $D_A(z)$, and $f(z)\sigma_8(z)$ from the
Sloan Digital Sky Survey DR7 Luminous Red Galaxies}, \emph{Mon. Not.
Roy. Astron. Soc.} {\bf 435(1)}
(2013) 255 
[arXiv:1209.0210].

\bibitem{Oka14}
A. Oka et al., \emph{Simultaneous constraints on the growth of
structure and cosmic expansion from the multipole power spectra of
the SDSS DR7 LRG sample},  \emph{Mon. Not. Roy. Astron. Soc.} {\bf
439(3)} (2014) 2515 
[arXiv:1310.2820].

\bibitem{BernalEtAl16}
  J.~L.~Bernal, L.~Verde and A.~G.~Riess,
  JCAP {\bf 1610} (2016) no.10,  019
  [arXiv:1607.05617 [astro-ph.CO]].

\bibitem{RiessEtAl19}
  A.~G.~Riess, S.~Casertano, W.~Yuan, L.~M.~Macri and D.~Scolnic,
  Astrophys.\ J.\  {\bf 876} (2019) no.1,  85
  [arXiv:1903.07603 [astro-ph.CO]].
  
\bibitem{pantheon}
  D.~M.~Scolnic {\it et al.},
  Astrophys.\ J.\  {\bf 859} (2018) no.2,  101
  [arXiv:1710.00845 [astro-ph.CO]].

\bibitem{Riess11}
Riess, A.G. {\it et al.},
Astrophys.\ J.\  {\bf 730} (2011) 119
Erratum: [Astrophys.\ J.\  {\bf 732} (2011) 129]
[arXiv:1103.2976 [astro-ph.CO]].

\bibitem{YuEtAl18}
  H.~Yu, B.~Ratra and F.~Y.~Wang,
  Astrophys.\ J.\  {\bf 856} (2018) no.1,  3
  [arXiv:1711.03437 [astro-ph.CO]].
  
\end{thebibliography}

\end{document}